\documentclass[twoside]{article}
\usepackage{fleqn,espcrc2}

\usepackage{graphicx}
\usepackage[figuresright]{rotating}

\newcommand{\AmS}{{\protect\the\textfont2
  A\kern-.1667em\lower.5ex\hbox{M}\kern-.125emS}}

\title{Anderson's ``Theorem'' and Bogoliubov-de Gennes Equations for
Surfaces and Impurities}

\author{K. Tanaka and F. Marsiglio,
\address{Department of Physics, University of Alberta,
Edmonton, Alberta, Canada T6G 2J1}}
       
\begin{document}

\begin{abstract}
In order to incorporate spatial inhomogeneity due to nonmagnetic
impurities, Anderson \cite{anderson59} proposed
a BCS-type theory in which single-particle states in such an inhomogeneous 
system are used.  We examine Anderson's proposal, in comparison with the 
Bogoliubov-de Gennes equations, for the attractive Hubbard model
on a system with surfaces and impurities.
\vspace{1pc}
\end{abstract}

\maketitle

The procedure for examining surface and impurity effects on a microscopic
level is by now well established. One uses a mean field-like decoupling,
with potentials which are determined from self-consistency requirements.
These potentials are then used in the effective Hamiltonian, which is
numerically diagonalized. This process is continued until self-consistency
is achieved. This is the essence of the Bogoliubov-de Gennes (BdG) 
\cite{deGennes66} formalism.

An earlier proposal was suggested by Anderson \cite{anderson59}, in 
which the single-particle problem is first diagonalized. Eigenvalues
and eigenstates are obtained, with which one can formulate the BCS
problem, but in a vector space associated with these eigenstates.
In certain
situations, the single-particle problem can be obtained analytically
(open boundaries, for example \cite{trugman99}), or it can be
obtained numerically with significantly less effort than required
by the full BdG process. In these instances it would be advantageous
to utilize the Anderson prescription. In this paper we report on
some test cases to evaluate the Anderson prescription. 

The BdG equations are well documented \cite{deGennes66}. In this work
we utilize the attractive Hubbard model, with open boundaries, and with
the possibility for single site impurity potentials. The resulting
equations are \cite{hirsch92}:
\begin{eqnarray}
E_n u_n(\ell) = \phantom{-} \sum_{\ell^\prime}
A_{\ell \ell^\prime}u_n(\ell^\prime) + \Delta_\ell v_n(\ell)
\label{bdgu}\\
E_n v_n(\ell) = -\sum_{\ell^\prime} A_{\ell \ell^\prime}v_n(\ell^\prime)
+ \Delta^\ast_\ell u_n(\ell)
\label{bdgv}
\end{eqnarray}
where
\begin{equation}
A_{\ell \ell^\prime} = -t \sum_\delta \biggl( 
\delta_{\ell^\prime, \ell - \delta} + \delta_{\ell^\prime, \ell + \delta}
\biggr)
+\delta_{\ell \ell^\prime} \biggl(V_\ell - \mu + \epsilon_\ell \biggr).
\label{bdgaux}
\end{equation}
The self-consistent potentials, $V_\ell$, and $\Delta_\ell$, are given
by
\begin{eqnarray}
\Delta_{\ell} = |U| \sum_n u_n(\ell) v_n^\ast(\ell) (1 - 2f_n)
\phantom{aaaaaaaaa}
\label{delpot} \\
V_\ell = -|U| \sum_n \biggl[
|u_n(\ell)|^2 f_n + |v_n(\ell)|^2 (1 - f_n) \biggr],
\label{harpot}
\end{eqnarray}
where $|U|$ is the strength of the attractive interaction, the index 
$n$ labels the eigenvalues (there are $2N$ of them), the index $\ell$
labels the sites (1 through N), and the composite eigenvector is
given by $\biggl({u_n \atop v_n} \biggr)$, of total length $2N$. The sums
in Eqs. (\ref{delpot},\ref{harpot}) are over positive eigenvalues only. 
The other
physical parameters are the single-particle hopping, $t$, the single site
impurity potentials, $\epsilon_\ell$, and the chemical potential, $\mu$.
The $f_n$ is the Fermi function, with argument $\beta E_n$, where
$\beta \equiv {1 \over k_BT}$, with $T$ the temperature. The single
site electron density, $n_\ell$, is given, through Eq. (\ref{harpot}),
by $V_\ell = -|U|{n_\ell \over 2}$. 

These equations are iterated to convergence, with results to be presented
below.

The alternative Anderson formalism \cite{anderson59} first solves 
for the eigenvalues and eigenstates of the `non-interacting' problem, i.e.,
\begin{equation}
E^0_n w_n(\ell) = \sum_{\ell^\prime}
A^0_{\ell \ell^\prime}w_n(\ell^\prime),
\label{andw}
\end{equation}
where
\begin{equation}
A^0_{\ell \ell^\prime} = -t \sum_\delta \biggl(
\delta_{\ell^\prime, \ell - \delta} + \delta_{\ell^\prime, \ell + \delta}
\biggr)
-\delta_{\ell \ell^\prime} \biggl(\mu - \epsilon_\ell \biggr).
\label{andaux}
\end{equation}
The $N\times N$ matrix equation (\ref{andw}) is solved for its
eigenvalues $E^0_n$ and eigenvectors $w_n$. This amounts to determining
the unitary matrix $U_{\ell n}$ that gives a basis for the electron
operators
\begin{equation}
c_{\ell \sigma}^\dagger = \sum_n U_{\ell n}^\ast \tilde{c}_{n \sigma}^\dagger,
\label{transform}
\end{equation}
which diagonalizes the single-particle Hamiltonian. From this matrix
we determine the transformed electron-electron interaction:
\begin{equation}
V_{nm,n^\prime m^\prime} = -|U| \sum_{\ell} U^\ast_{\ell n} U^\ast_{\ell m}
U_{\ell n^\prime} U_{\ell m^\prime},
\label{interaction}
\end{equation}
which now mediates the (generally off-diagonal) electron-electron interaction.
The gap and number equations are derived in the usual way; they are in
general complicated --- the gap is a function of the quantum label $n$ and
the chemical potential is shifted by an $n$-dependent quantity. Once these
are obtained, we can transform back to real space, and examine the
gap function or the electron density, for example, as a function of
position.

Figure 1 illustrates the gap parameter obtained by the BdG formalism
as a function of position, for all
densities, in the case of open boundary conditions (OBC), in one dimension.
Results in higher dimension will be very similar \cite{tanaka00}.
Variations in the gap are strongest near the boundaries, as expected, 
and the Anderson prescription is reasonably accurate in
reproducing the oscillations (not shown). 
In Fig. 2 we show the gap as a function
of position for the case of a single impurity (at site 16) with a repulsive 
potential, with periodic boundary conditions (PBC).
As expected, the gap is suppressed at this site, and once again,
the Anderson prescription semi-quantitatively reproduces the BdG result.
\begin{figure}[htb]
\begin{center}
\includegraphics[width=17pc,height=13pc]{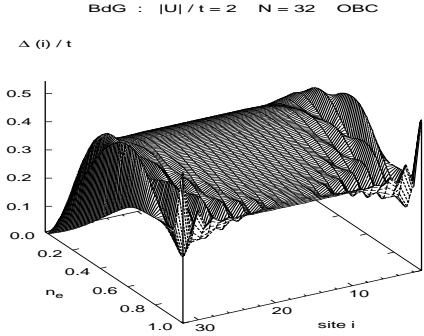}
\caption{$\Delta(i)$ for all densities $n_e$ (OBC).}
\label{fig1}
\end{center}
\end{figure}
\begin{figure}[htb]
\begin{center}
\includegraphics[width=15pc,height=9pc]{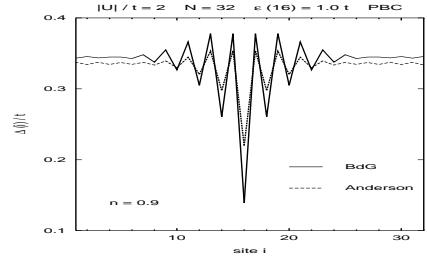}
\caption{$\Delta(i)$ for $n_e=0.9$ ($\epsilon(16)=1.0\,t$).}
\label{fig2}
\end{center}
\end{figure}

\end{document}